\documentclass[12pt]{iopart}
\usepackage{graphicx}

\begin{document}

\title[]{Proton induced reaction on $^{108}$Cd for astrophysical p-process studies}

\author{Sukhendu Saha$^{1,3}$, Dipali Basak$^{1,3}$, Tanmoy Bar$^{1,3}$, Lalit Kumar Sahoo$^{1,3}$, Jagannath Datta$^{2}$, Sandipan Dasgupta$^{2}$, Norikazu Kinoshita$^{4}$, and Chinmay Basu$^{1,3}$}

\address{$^1$Saha Institute of Nuclear Physics, 1/AF, Bidhannagar, Kolkata$-$700064, India}
\address{$^2$Analytical Chemistry Division, Bhabha Atomic Research Centre, Variable Energy Cyclotron Centre, 1/AF Bidhannagar, Kolkata 700064, India}
\address{$^3$Homi Bhabha National Institute, Mumbai, Maharashtra$-$400094, India}
\address{$^4$Institute of Technology, Shimizu Corporation, Koto-ku, Tokyo 135-8530, Japan}

\ead{sukhendu.saha@saha.ac.in}
\vspace{10pt}
\begin{indented}
\item[]
\end{indented}

\begin{abstract}
The proton capture cross-section of the least abundant proton-rich stable isotope of cadmium, $^{108}$Cd (abundance 0.89\%), has been measured near the Gamow window corresponding to a temperature range of 3-4 GK. The measurement of the 
$^{108}$Cd(p,$\gamma$)$^{109}$In reaction was carried out using the activation technique. The cross-section at the lowest energy point of 3T$_9$, E$_p$$^{lab}$= 2.28 MeV, has been reported for the first time. The astrophysical S-factor was measured in the energy range relevant to the astrophysical p-process, between 
E$_p$$^{cm}$= 2.29 and 6.79 MeV. The experimental results have been compared with theoretical predictions of Hauser-Feshbach statistical model calculations using TALYS-1.96. A calculated proton-optical potential was implemented to achieve better fitting, with different combinations of available nuclear level densities (NLDs) and $\gamma$-ray strength functions in TALYS-1.96. The calculations provided satisfactory agreement with the experimental results. The reaction rate was calculated using the calculated potential in TALYS-1.96 and compared with the values provided in the REACLIB database.
\end{abstract}

%
\vspace{2pc}
%
\submitto{\JPG}
%
%
%

\section{Introduction}
\label{int}
\begin{figure}[h]
\begin{center}
\includegraphics[scale=0.5]{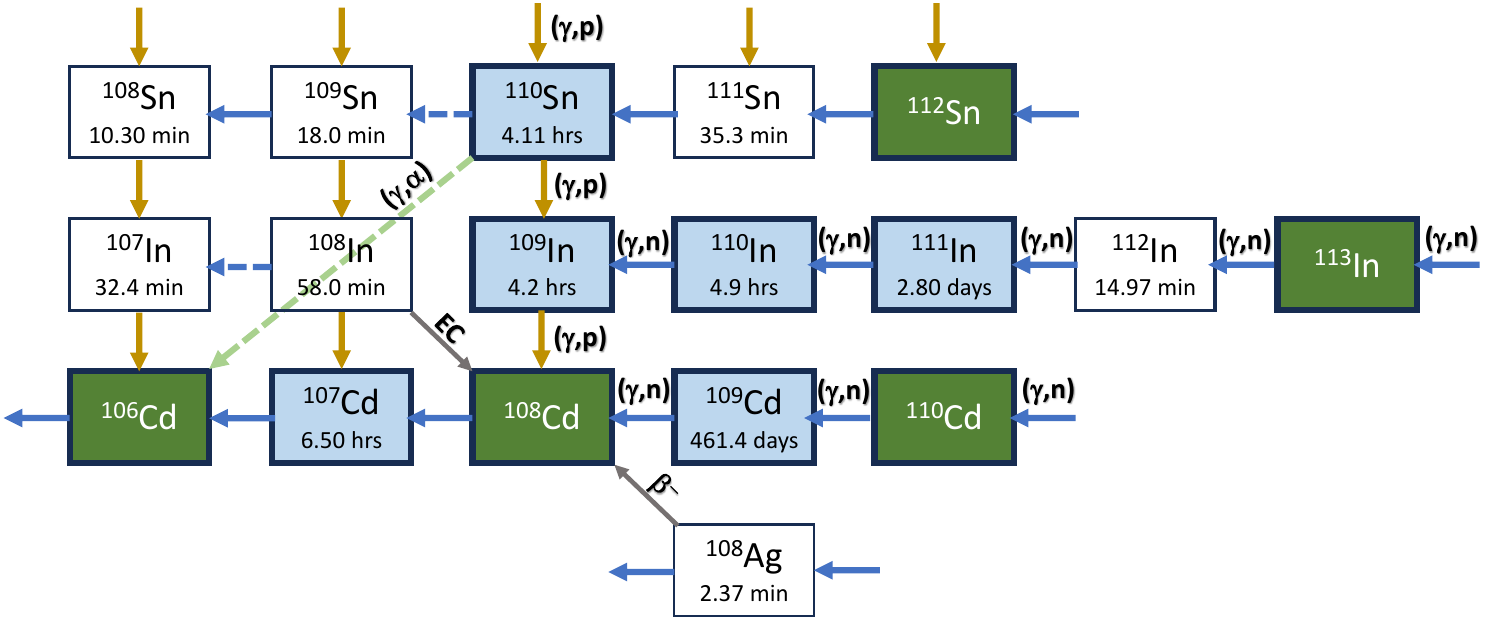}
\end{center}
\caption{Dominant reaction network for the production of $^{108}$Cd at 3 GK. The sequence of ($\gamma$,n) reactions starting from stable nuclei is illustrated. Solid blue arrows indicate dominant ($\gamma$,n) reactions, while orange arrows represent ($\gamma$,p) reactions. Gray arrows denote weak interactions, and dotted arrows indicate less dominant reactions \cite{rauscher2013constraining}.}\label{Fig1} 
\end{figure}

There are about 34 neutron deficient p-nuclei, that are produced neither by the s or r process directly. The most probable synthesis mechanism of these nuclei in the supernovae are through $\gamma$-process that involves ($\gamma$,n), ($\gamma$,p) and ($\gamma$,$\alpha$) reactions. At around 2-3 GK temperature, the photon density is high and photo-nuclear reactions play an important role in nucleosynthesis \cite{arnould2003p, woosley1978p, PhysRevC.73.015804, rauscher2013constraining}. The $^{108}$Cd is one such p-nucleus whose synthesis is important but comparatively less studied in the literature \cite{gyurky2007proton, olivas2020measurements, skakun1992level, roughton1979thick}. Its formation path can be either from $^{109}$Cd through ($\gamma$,n) reaction or from $^{109}$In through ($\gamma$,p) reaction (shown in Figure~\ref{Fig1}). Due to difficulty in mono-energetic photon beams with a radioactive target, usually the $\gamma$-process reactions have been studied through their inverse counterparts \cite{gyurky2019activation}. The $^{108}$Cd(n,$\gamma$) reaction, which is the inverse of $^{109}$Cd($\gamma$,n) reaction, has been studied extensively in the literature \cite{de1978neutron, gicking2019neutron, anufriev1984parameters, beda1964cross, mangal1962thermal, boyd1964thermal}. Comparatively, $^{108}$Cd(p,$\gamma$) has been less studied. The most recent measurement \cite{olivas2020measurements} uses an advanced sum-spectrometer to measure $^{108}$Cd(p,$\gamma$) cross-sections using in-beam technique. The measurement are for the proton energies from 3.5 to 7.0 MeV. However, the target used in their experiment is very thick (2.09 mg/cm$^2$) and characterisation of the target is not available. Gyurky et al. \cite{gyurky2007proton} on the other hand uses the activation method and measures the $^{108}$Cd(p,$\gamma$) cross-section between E$_p$ = 2.4 -- 4.8 MeV and using thinner targets (100 -- 500 $\mu$g/cm$^2$). However, their target enrichment is only 2.05\%.

In this work, we present measurement of $^{108}$Cd(p,$\gamma$) cross-section in the energy range E$_p$ = 2.28 -- 6.85 MeV using a higher energy accelerator. In order to measure the cross-sections at low energies the stack foil activation technique is used. The determining of the reaction cross-section, establishes that even in the absence of a low-energy, high-current accelerator, the stack foil activation technique can effectively be used to measure astrophysical cross-sections. A 66.3\% enriched $^{108}$Cd target was prepared and used in the present work. This is a betterment of the measurements in comparison to that of Gryurky et. al. \cite{rauscher2013constraining}, whose target enrichment was only 2.05\%. A Gaussian averaged cross-section correction is determined to account for the inherent energy uncertainty of the Cyclotron beam and energy loss and straggling due to the use of stacked targets.
A proton optical potential is suggested from systematics using published proton elastic scattering data from p-nuclei and isotopes with masses near those of p-nuclei. This potential predicts a better astrophysical S-factor compared to the well known proton potentials in the literature.

\section{Experimental Procedure}
\label{sec2}
The $^{108}$Cd(p,$\gamma$)$^{109}$In cross-section measurement was performed at the K130 Cyclotron, Variable Energy Cyclotron Centre (VECC), Kolkata, India, using the activation technique. The cross-section was measured at nine energy points ranging from 2.29 MeV to 6.85 MeV. Four stack setups were individually irradiated with a 7 MeV proton beam for 6 to 17 hours, depending on the theoretically estimated yield, with a beam current of 150-200 nA. Aluminum foils with thicknesses of 6.5 $\mu$m, 12 $\mu$m, 25 $\mu$m, and 100 $\mu$m, and a purity of 99.95\%, were used as degrader foils in the stack setups. The irradiated targets were cooled and then placed for offline gamma measurement. The detailed experimental procedure is described below.

\subsection*{Target preparation}
\label{sec2a}
The 66.3$\%$ enriched $^{108}$Cd targets on mylar (H$_8$C$_{10}$O$_4$) backing were prepared using the vacuum evaporation technique. The evaporation setup was optimized by adjusting parameters such as the distance between the crucible and substrate holder, the dimensions of the crucible opening, the \textit{e$^-$}-beam current, and the evaporation time. A total of 38.7 mg of enriched $^{108}$Cd was utilized with an \textit{e$^-$}-beam current ranging from 3 to 7 mA. The crucible opening was set at 4 mm, and the distance between the crucible and substrate holder was 5 cm. The thickness of the cadmium targets was determined through $\alpha$-energy loss measurements and validated by RBS measurement. X-ray Photoelectron Spectroscopy (XPS) and X-ray Fluorescence (XRF) analyses were performed to check for elemental contamination, as well as to assess surface morphology and chemical composition. The experiment was conducted using eight different targets with thicknesses ranging from 288 $\mu$g/cm$^2$ to 659 $\mu$g/cm$^2$ on a mylar backing. The uncertainty in target thickness is about 10$\%$.

\begin{table}[h]
\label{tab1}
\caption{ Thickness of Cd targets used during the $^{108}$Cd(p,$\gamma$)$^{109}$In experiment. Thickness is represented in $\mu$g/cm$^2$. Energy* is the mean energy of proton beam falling on the target in MeV.}
\resizebox{\textwidth}{!}{%
    \begin{tabular}{c p{2.5cm} p{2.5cm} p{2.5cm} p{2.5cm} p{2.5cm} p{2.5cm} p{2.5cm} p{2.5cm} p{2.5cm} p{2.5cm}}
    \hline 
    & Target name & Cd1 & Cd2 & Cd3 & Cd4 & Cd5 & Cd6 & Cd7 & Cd8 & Cd9 \\
    \hline
    & Thickness & 288(29) & 311(31) & 323(32) & 506(51) & 597(60) & 642(64) & 659(66) & 656(66) & 288(29) \\
    & ($\mu$g/cm$^2$) & \\
    \hline
    & Energy* & 6.85(3) & 6.10(4) & 5.21(6) & 4.78(7) & 4.05(8) & 3.63(9) & 2.98(9) & 2.28(12) & 6.69(3) \\
    & (MeV) &\\
    \hline
    \end{tabular}}
\end{table}

\subsection*{Target stacks and irradiation setup}
\label{sec2b}

\begin{figure}[h]
\begin{center}
\includegraphics[scale=0.4]{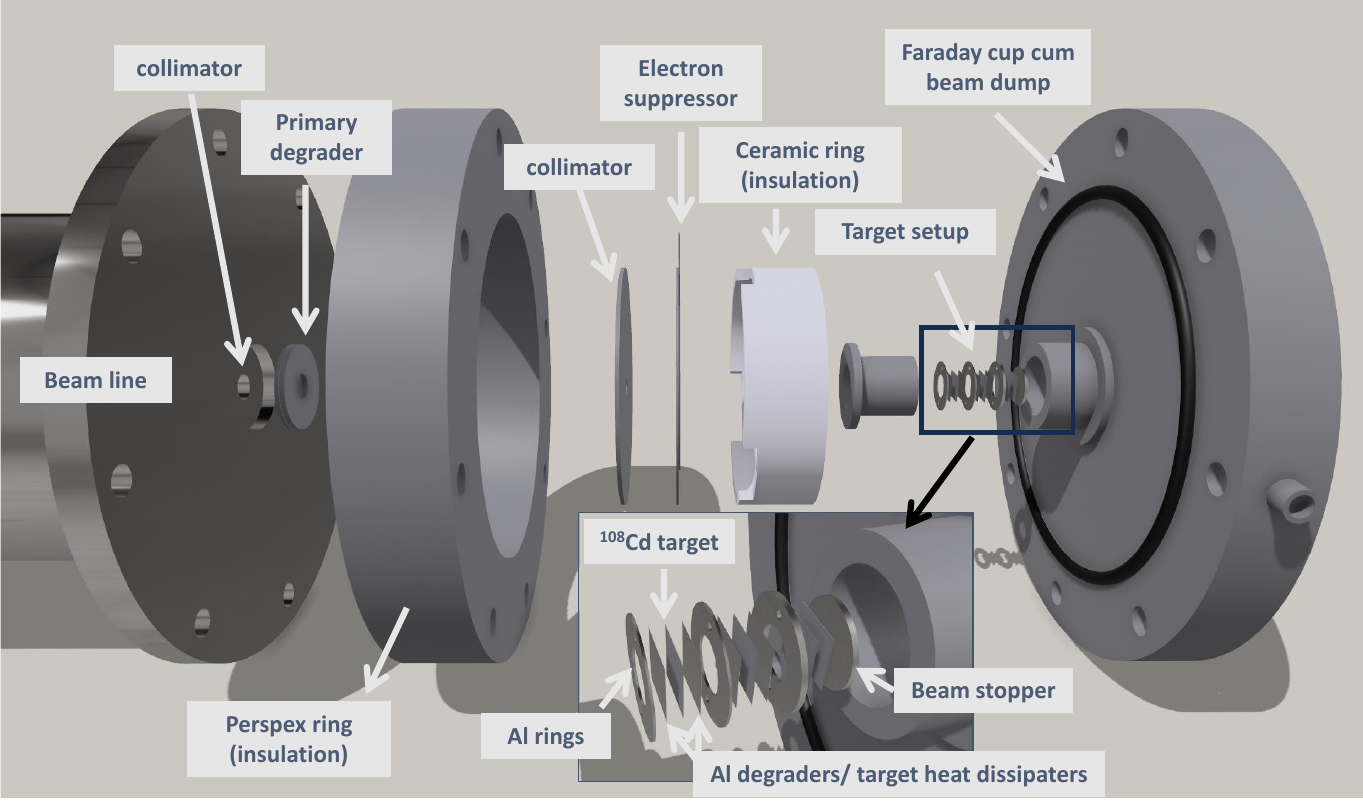}
\end{center}
\caption{Schematic of the irradiation setup. A target-stack setup at the beam line of K130 Cyclotron for $^{108}$Cd(p,$\gamma$)$^{109}$In measurement.}\label{Fig2} 
\end{figure}

The $^{108}$Cd(p,$\gamma$) reaction cross-section was measured offline using the activation technique. Four different stack setups were individually irradiated with a proton beam. The 7 MeV proton beam from the K130 Cyclotron was primarily degraded using a series of Al foils of specific thicknesses in the primary degrader arrangement. The degraded proton beam was then focused through a 6 mm diameter collimator before reaching the target stack setup. An electron suppressor with a suppression voltage of --300 V was used during the irradiation process. A ceramic ring was employed to isolate the end flange (Faraday cup) from the negative high voltage suppression ring, while a perspex ring provided electrical isolation between the end flange and the beam line (Figure~\ref{Fig2}). 

\begin{figure}[h]
\begin{center}
\begin{tabular}{cc}
\includegraphics[scale=0.3]{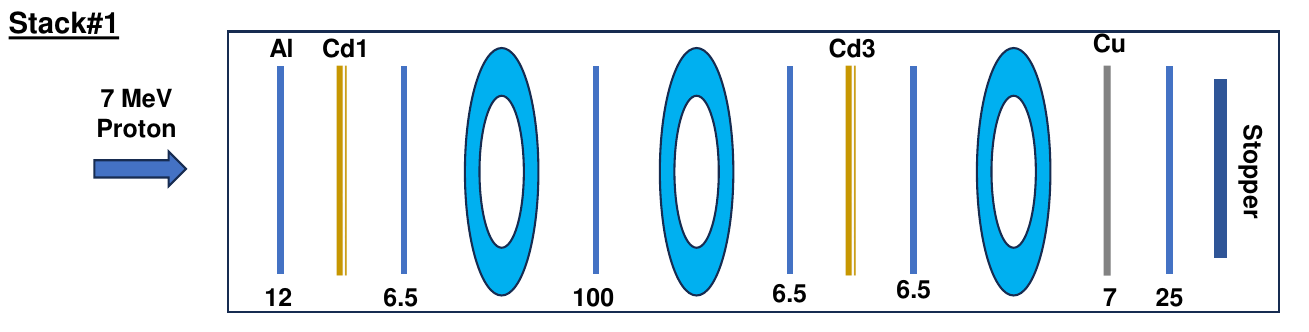}
\includegraphics[scale=0.3]{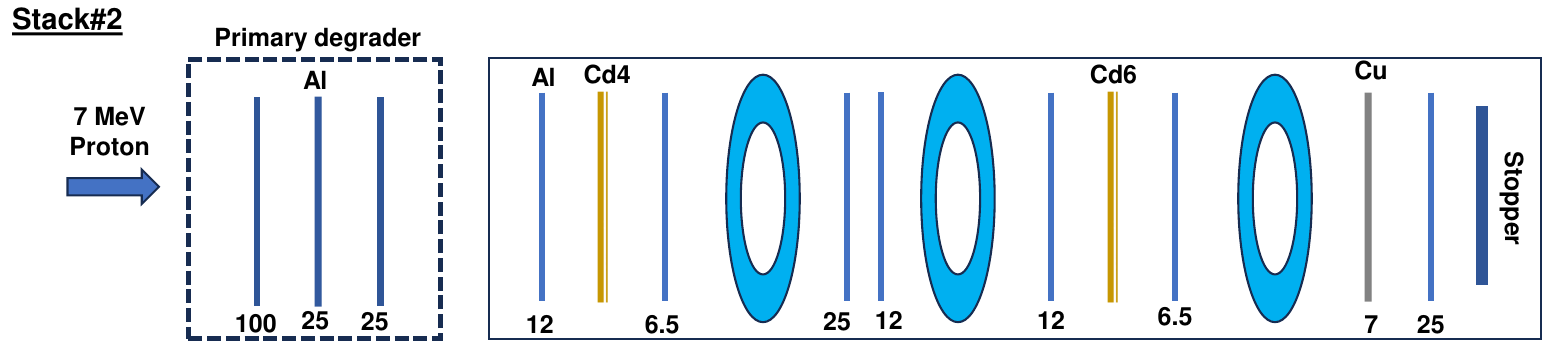}\\
\includegraphics[scale=0.3]{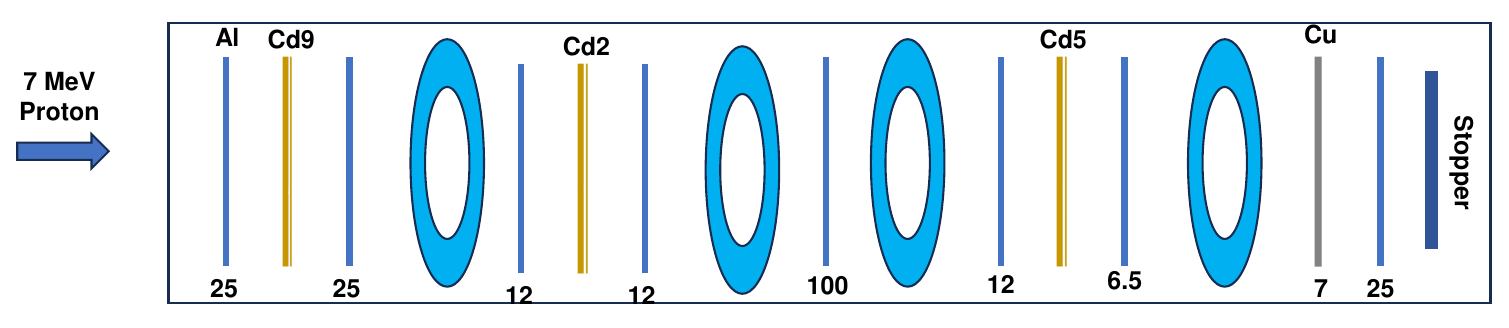}
\includegraphics[scale=0.3]{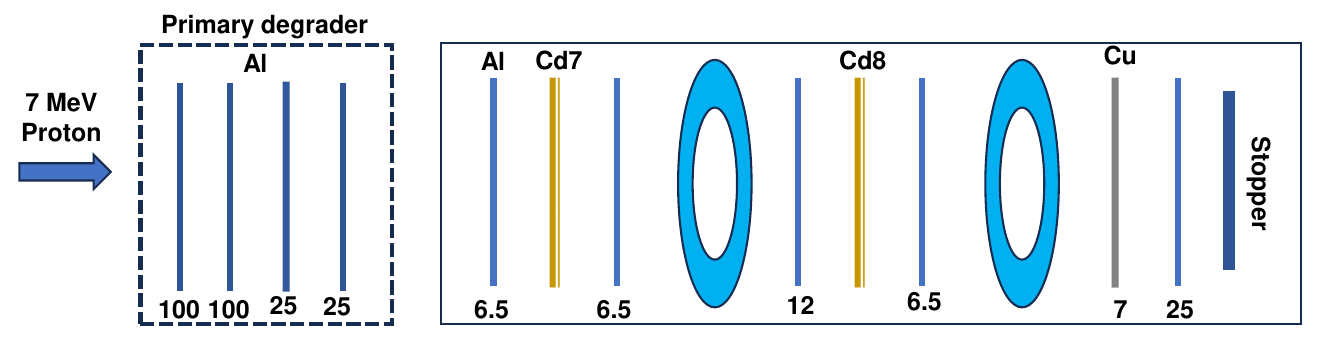}\\
\end{tabular}
\end{center}
\caption{Four different stack-foil setups, Stack\#1 to Stack\#4 shown respectively. Blue lines indicating Al foils of different thicknesses (numbers in the schematics are in $\mu$m). Al rings were used to hold the targets straight. Cu foils were used to monitor the beam current.}\label{Fig3}
\end{figure}

Stack\#1, Stack\#2, and Stack\#4 each consisted of two targets, while Stack\#3 contained three targets, shown in Figure~\ref{Fig3}. They were irradiated for 6.13 hours, 13.47 hours, 16.56 hours, and 8.22 hours, respectively. The mean proton energies incident on the targets were 6.85 MeV \& 5.21 MeV, 4.78 MeV \& 3.63 MeV, 3.02 MeV \& 2.31 MeV, and 6.69 MeV, 6.10 MeV \& 4.05 MeV, respectively. The beam current ranged from 160 to 200 nA. Due to fluctuations, the beam current was digitally recorded from the Faraday cup and accounted for during the analysis. The $^{108}$Cd on mylar backing targets were sandwiched with Al foils, which acted as both energy degraders and heat dissipators for the target foils. During the irradiation of Stack\#1 and Stack\#3, the primary degrader setup was not used. Al rings were placed to hold the targets straight and tight in place. The end flange with the target stack was cooled using low-conductivity water.

\subsection*{Offline $\gamma$-ray measurement}
\label{sec2d}
After irradiation, each target stack was cooled for 1-2 hours to reduce the unnecessary $\gamma$-peaks from short-lived states. The targets were then individually mounted on a plastic sheet, and $\gamma$-photons were counted. The $\gamma$-activity was measured using a CANBERRA high-purity germanium (HPGe) detector with a 40\% relative detection efficiency and an energy resolution of 1.8 keV at 1.33 MeV of $\gamma$ energy.

The targets were positioned at optimal distances (12.5 mm, 37.5 mm, and 50.5 mm) in front of the detector to minimize detection dead time while ensuring sufficient photo-peak counts. The counting duration for the targets ranged from 15 minutes to 3.5 hours, depending on their activity. During $\gamma$-ray counting, both the target and the detector were shielded by 7.5 cm thick lead bricks. Data acquisition was performed using a 16K channel integrated MCA based on digital signal processing technology, CANBERRA DSA1000, with GENIE spectroscopy software.

\subsection*{Detector calibration}
\label{sec2e}
A $^{125}$Eu point source with a known DPS (A$_\circ$ = 3.908 × 10$^4$ Bq on \textit{17$^{th}$ May, 1982}) was used to calibrate the detector and measure the absolute efficiency of the photo-peak at different energy points. The $^{125}$Eu source, with known $\gamma$-ray energies of 121.8 keV, 344.3 keV, 778.9 keV, 964.1 keV, 1112.1 keV, and 1408.0 keV, was employed for detector calibration. To measure the detector efficiency, the activity of $^{125}$Eu was measured at all three positions where the $\gamma$-spectra were counted for the irradiated targets, under the same detector and target shielding conditions.

\section{Data Analysis and Results}

\subsection{Experimental cross-section}
\label{sec3a}

\begin{figure}[!h]
\begin{center} 
\includegraphics[scale=0.5]{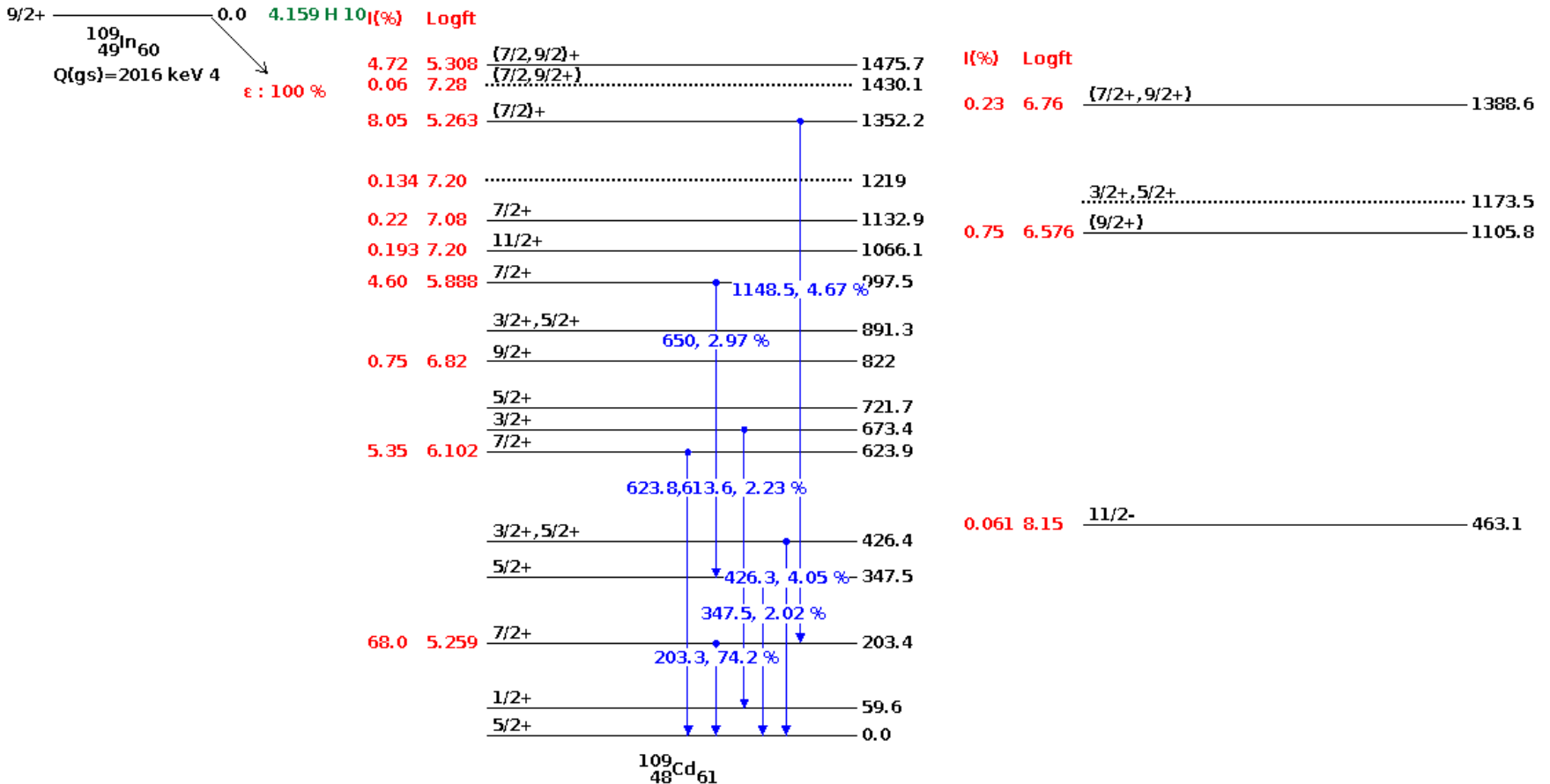}
\end{center}
\caption{Decay scheme of $^{109}$In to $^{109}$Cd. Emission of intense $\gamma$-rays (relative intensity $>$2\%) followed by electron-capture. Figure generated from \cite{NNDC_NuDat3}.}\label{Fig4}
\end{figure}

The produced nuclei, $^{109}$In, from the $^{108}$Cd(p,$\gamma$) reaction decay to $^{109}$Cd via electron capture, with a half-life of t$_{1/2}$ = 4.159 hours. Characteristic gamma-rays are emitted as the excited states of $^{109}$Cd de-excite to the ground state. The decay scheme is shown in Figure~\ref{Fig4}. The most intense gamma-rays counted were 203.3 keV (relative intensity 74.2\%), decaying from the 203.4 keV (7/2$^+$) state to the ground state (5/2$^+$), and 623.8 keV (relative intensity 5.64\%), decaying from the 623.9 keV (7/2$^+$) state to the ground state of $^{109}$Cd.

The total number of decays of the residual nuclei in the counting period can be estimated by given relation,

\begin{equation}
\centering
N_{decay} = \frac{A_\gamma}{\epsilon I_\gamma}
\end{equation}

To determine the peak count (A$_\gamma$) in the gamma spectrum for a specific transition, the background counts were subtracted from the adjacent higher and lower energy sides using the CERN-ROOT data analysis tool. Here, $\epsilon$ represents the detector efficiency, and I$_\gamma$ denotes the intensity of the gamma transition.

The reaction cross-section at different mean proton energy can be determined by,
\begin{equation}
  \sigma_{reac}=\frac{\lambda A_\gamma}{\epsilon I_\gamma N_A \phi_b (1-e^{-\lambda t_{irr}})e^{-\lambda t_{wait}}(1-e^{-\lambda t_{meas}})} 
\end{equation}

\begin{figure}[h!]
\begin{center} 
\includegraphics[scale=0.1]{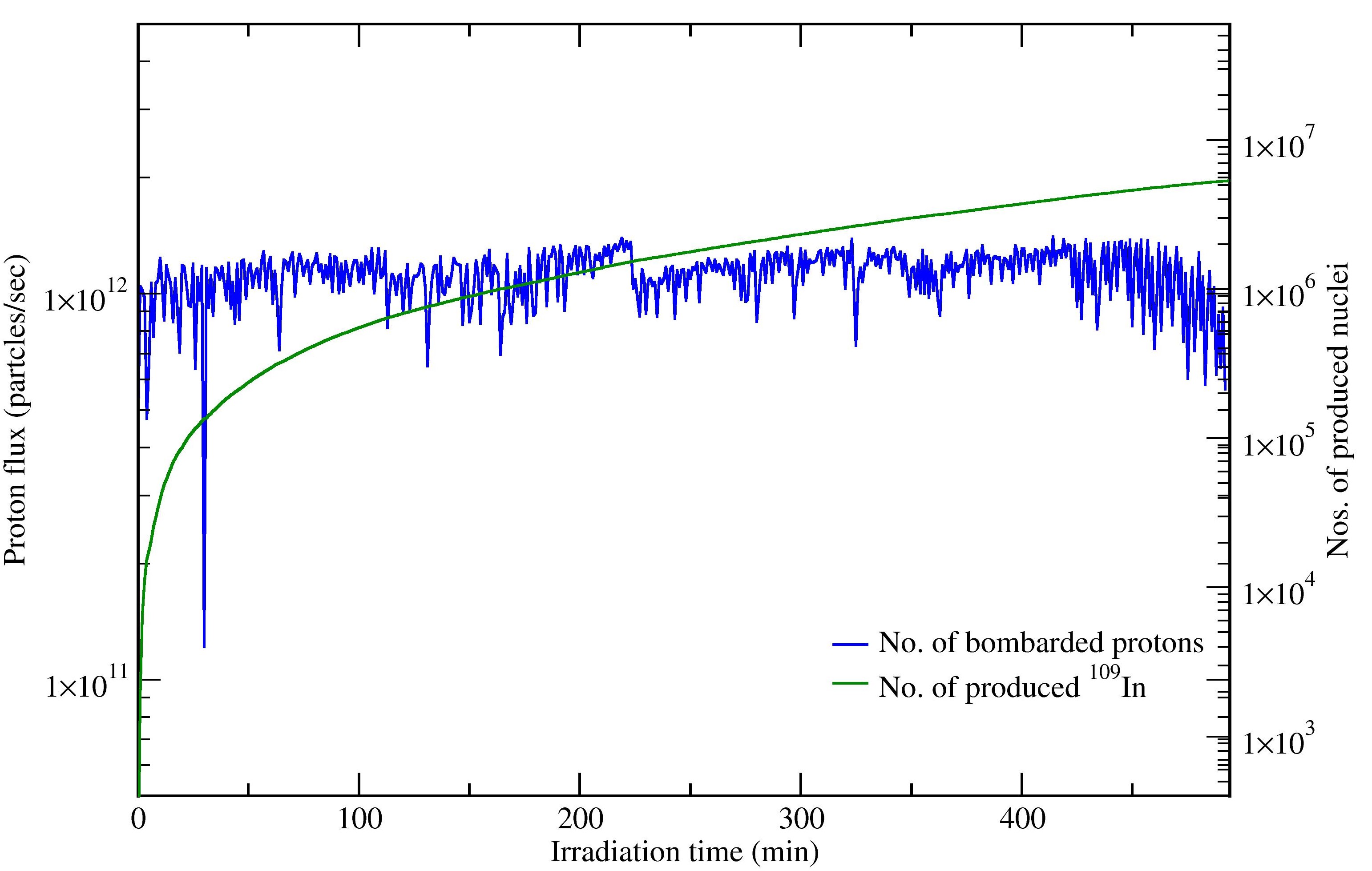}
\end{center}
\caption{Proton flux with time and number of produced residue nuclei for 5.20 $\pm$ 0.06 MeV of proton beam on $^{108}$Cd.}\label{Fig5}
\end{figure}

Where $\lambda$ is the decay constant of the residual nuclei ($\lambda = ln (2)/t_{1/2}$). A$_\gamma$ is the peak area under the 203.3 keV gamma-peak, t$_{irr}$, t$_{wait}$ and t$_{meas}$ represent the irradiation time, waiting time, and measurement time, respectively. $\phi _b$ is the proton beam flux (particles/sec). Since the beam current fluctuates during the irradiation period (as shown in figure~\ref{Fig5}), the irradiation time was divided into N intervals of $\Delta$t = 1 minute. During each $\Delta$t interval, the beam flux at the i$^{th}$ interval $\phi_{bi}$ is assumed to be constant. The cross-section formula was then modified accordingly,
\begin{equation}
  \sigma_{reac}=\frac{\lambda A_\gamma}{\epsilon I_\gamma N_A e^{-\lambda t_{wait}}(1-e^{-\lambda t_{meas}}) \sum_{i=1}^{N} \phi_{bi} \frac{1-e^{-\lambda \Delta t}}{\lambda} e^{-\lambda \Delta t (N-i)}}
\end{equation}

\subsection*{Energy and cross-section correction in the stack foil activation cross-section measurements}
\label{sec3aa}

\begin{figure}[h]
\begin{center}
\includegraphics[scale=0.3]{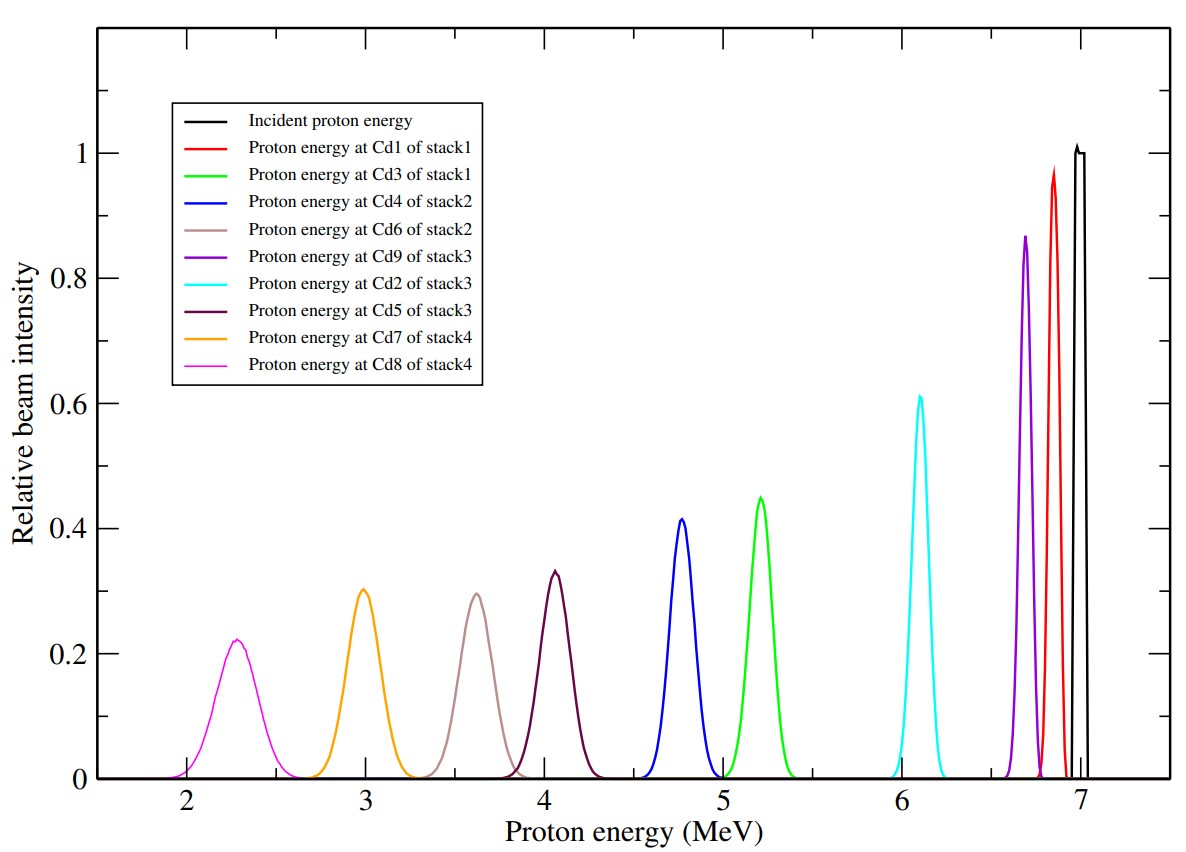}
\end{center}
\caption{PHITS simulation spectrum of proton energy point at each target position}\label{Fig6}
\end{figure}

\begin{table}[h!]
  \centering
  \caption{$^{108}$Cd(p,$\gamma$)$^{109}$In reaction cross-section and experimental S-factor.}\label{tab2}
   \begin{tabular}{lllll} 
    \hline
        \textbf{Effective } & \textbf{$\tilde{E}_{CM}$ } &  \textbf{Experimental} & \textbf{Corrected} & \textbf{S-factor} \\
        \textbf{beam-energy} & \textbf{(MeV)} & \textbf {cross-section } & \textbf{cross-section } & \textbf{($\times 10^{7}$ MeV-b)} \\
         \textbf{$\tilde{E} _{Lab}$ (MeV)}& &\textbf{($\mu$b)}&\textbf{($\mu$b)}&\\
     \hline   
    
      2.26 $\pm$ 0.12 & 2.24 $\pm$ 0.12 & 1.4 $\pm$ 0.2 & 1.4 $\pm$ 0.2 & 11.72 $\pm$ 2.12 \\
      2.96 $\pm$ 0.09 & 2.93 $\pm$ 0.09 & 83.4 $\pm$ 8.4 & 83.3$\pm$	8.4& 18.15 $\pm$ 2.55\\
      3.60 $\pm$ 0.09 & 3.57 $\pm$ 0.09&  527.6 $\pm$ 55.5 & 527.4	$\pm$ 55.5 & 11.05 $\pm$ 1.52\\
      4.05 $\pm$ 0.08 & 4.01 $\pm$ 0.08& 1451.6 $\pm$ 146.2 &1451.3 $\pm$ 146.2 & 10.20 $\pm$ 1.34\\
      4.76 $\pm$ 0.07 & 4.72 $\pm$ 0.07 & 4232.9 $\pm$ 431.9 & 4232.8 $\pm$ 431.9 & 5.39 $\pm$ 0.64 \\
      5.20 $\pm$ 0.06 & 5.15 $\pm$ 0.06 & 11508.1 $\pm$ 1158.1 & 11508.4 $\pm$	1158.2 & 6.39 $\pm$ 0.74\\
      6.09 $\pm$ 0.04 & 6.03 $\pm$ 0.04 & 36456.9 $\pm$ 3672.7 & 36458.1 $\pm$ 3672.8 & 4.90 $\pm$ 0.55\\
      6.68 $\pm$ 0.03 & 6.62 $\pm$ 0.03 & 27778.1 $\pm$ 2794.6 & 27778.0 $\pm$ 2794.6 & 1.72 $\pm$ 0.18\\
      6.84 $\pm$ 0.03 & 6.78 $\pm$ 0.03 & 40433.2 $\pm$ 4071.4 & 40433.0 $\pm$ 4071.4 & 2.06 $\pm$ 0.74\\
      
     \hline  
\end{tabular}
\end{table}

In the measurement of nuclear cross-sections using the stack foil activation method, corrections to the beam energy and cross-section are essential due to significant energy loss and straggling of the beam within the stacked target setup. Since the cross-section is derived from the activity of the irradiated target, and the yield does not depend explicitly on the energy, it is important to correct for the variation of yield due to the variation in beam energy.

As described \cite{kinoshita2016proton, brune2013energy}, the average beam energy at each Cd target is determined by modeling the energy distribution as a Gaussian, where the width of the distribution represents the energy straggling. The mean energy loss of the beam is calculated using SRIM \cite{ziegler2010md}, while the energy straggling is evaluated with the PHITS code \cite{sato2013particle, kinoshita2016proton}. The mean energy at which the cross-section is reported is determined by the following method:

\begin{equation}
   \tilde{E}=\frac{\int_{E_0 - \Delta}^{E_0} E \sigma_{th}(E)f(E) \,dx}{\int_{E_0 - \Delta}^{E_0} \sigma_{th}(E)f(E) \,dx}
\end{equation}

Where \textit{E$_0$} is the beam energy before the beam falling on the target and \textit{$\Delta$} is the beam energy loss through the target. The Gaussian distribution of the beam is \textit{f(E)},

\begin{equation}
    f(E) = \frac{1}{\sqrt{2\pi \sigma^2}} e^{-\frac{(E - E_x)^2}{2\sigma^2}}
\end{equation}

Where \textit{E$_x$} (determined from SRIM) is the mean energy of the beam falling on the target and $\sigma$ is its width (determined from PHITS).
Having obtained the mean energy, the cross-section evaluated at \textit{E$_0$} is corrected by multiplying a factor \textit{c}, given by,

\begin{equation}
    c = \frac{\int_{E_0 - \Delta}^{E_0}\sigma_{th}(E)f(E) \,dx}{\sigma_{th}(\tilde{E})\int_{E_0 - \Delta}^{E_0} f(E) \,dx}
\end{equation}

The PHITS \cite{sato2013particle} calculations for energy straggling at various positions of all the stacks are shown in figure~\ref{Fig6}. 

\begin{figure}[h]
\begin{center}
\includegraphics[scale=0.3]{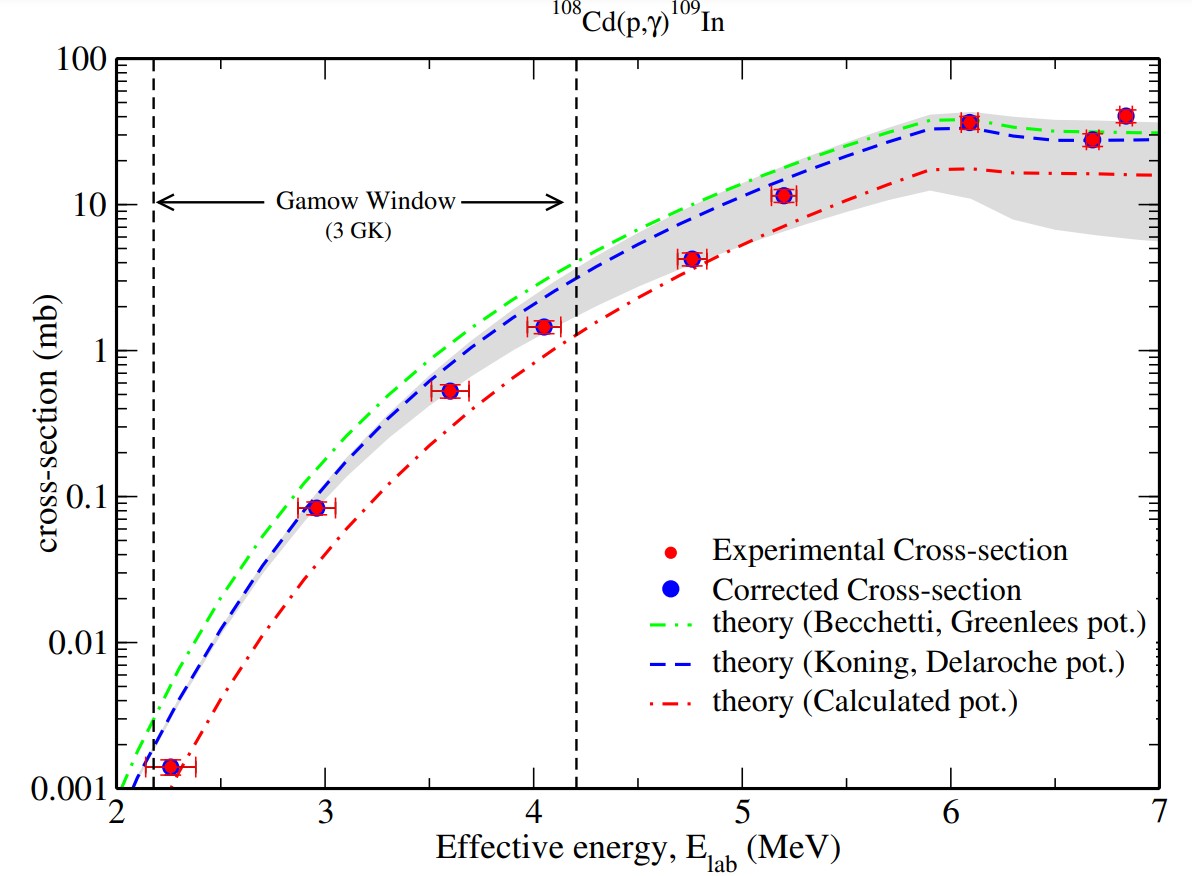}
\end{center}
\caption{Measured $^{108}$Cd(p,$\gamma$)$^{109}$In reaction cross-section and corrected cross-section (correction factor for energy discrepancy due to stacked target setup). Theoretically reproduced reaction cross-section by varying input parameters (Koning potential \cite{koning2003local}, SHFB, BAL; Becchetti et. al potential \cite{becchetti1969nucleon}, SHFB, BAL and Calculated potential, SHFB, BAL) using TALYS-1.96. Gray-shaded area represents the range of theoretical data reproducible in TALYS-1.96 using various nuclear input parameter combinations (p-OMP, NLDs, and $\gamma-$SFs)}\label{Fig7}
\end{figure}

\begin{figure}[h]
\begin{center}
\includegraphics[scale=0.3]{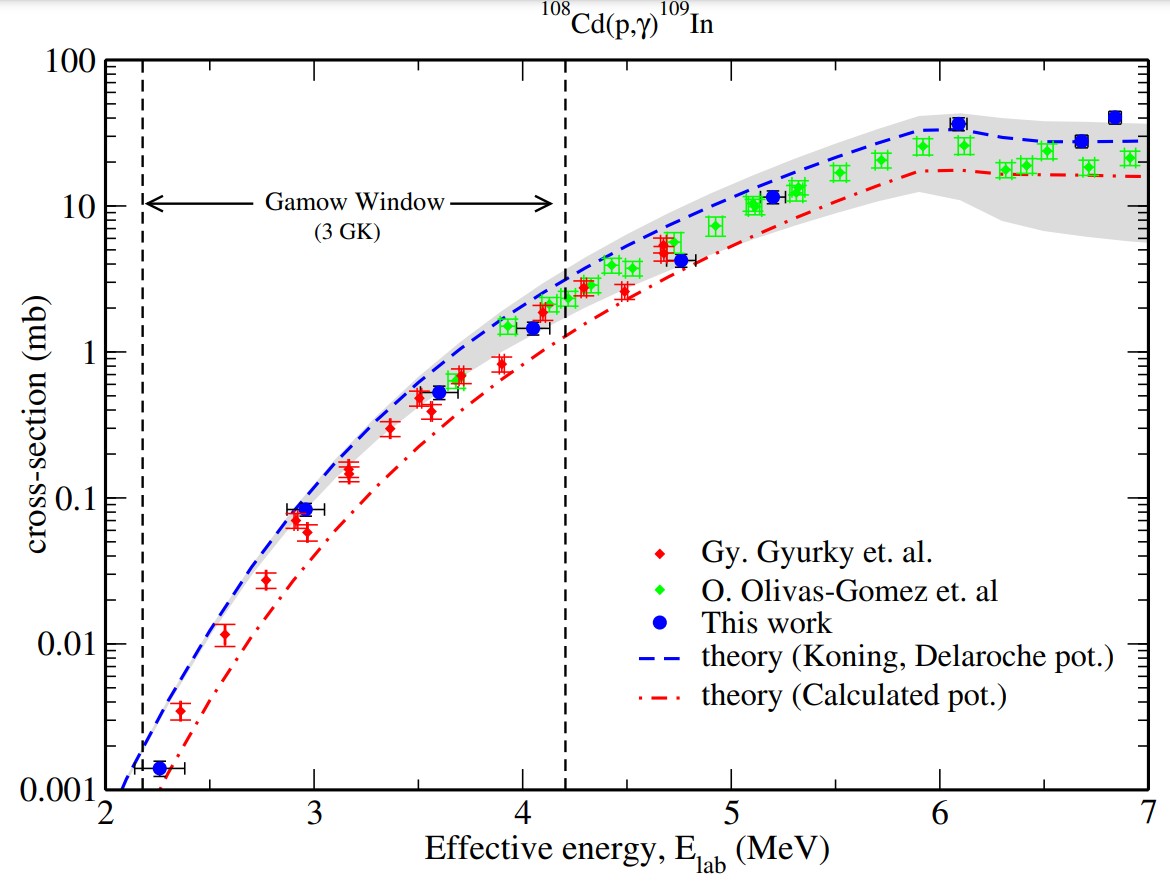}
\end{center}
\caption{Same as Figure~\ref{Fig7}. Previously measured data for $^{108}$Cd(p,$\gamma$) shows no discrepancies \cite{gyurky2007proton, olivas2020measurements}. Data retrieved from \cite{IAEA_EXFOR}. }\label{fig8}
\end{figure}

The theoretical cross-sections, $\sigma_{th}(E)$ and $\sigma_{th}(\tilde{E})$, are calculated at energy $E$ and effective energy $\tilde{E}$ using the statistical model code TALYS-1.96 with default input parameters. The effective energy $\tilde{E}$ (from equation 4) and correction factor $c$ (from equation 6) were computed using Python. The experimentally measured cross-sections, along with the corrected cross-sections at the effective lab energy, are listed in Table~\ref{tab2} and depicted in Figure~\ref{Fig7}. The 1-$\sigma$ uncertainty from PHITS (Figure~\ref{Fig6}) is taken as the energy error.

\subsection{Theoretical prediction and Proton optical potential for p-nuclei}
\label{}
The experimental cross-section for the $^{108}$Cd(p,$\gamma$)$^{109}$In reaction was reproduced using the Hauser-Feshbach statistical model code TALYS-1.96 \cite{koning2023talys, IAEA_TALYS_Tutorial}. Theoretical cross-section predictions in TALYS involved testing 216 different combinations of nuclear input parameters \cite{bar2024measurement}, including proton-nucleus optical potential models (p-OMPs), nuclear level densities (NLDs), and gamma-ray strength functions ($\gamma$-SFs).
TALYS-1.96 provides four types of proton optical model potentials (p-OMPs) \cite{IAEA_TALYS_Tutorial}, and in addition, the Becchetti-Greenlees potential \cite{becchetti1969nucleon} and a calculated optical model potential was included to better match the experimental data.

In Figure~\ref{Fig7}, the gray-shaded area represents the range of theoretical data reproducible in TALYS-1.96 using various nuclear input parameter combinations (p-OMP, NLDs, and $\gamma-$SF). It was observed that the theoretical cross-section calculated with the Koning potential (TALYS default) \cite{IAEA_TALYS_Tutorial, koning2003local}, combined with Skyrme-Hartree-Fock-Bogolyubov level density (microscopic model) \cite{goriely2006microscopic} and the Brink-Axel-Lorentzian (BAL) \cite{axel1962electric, brink1957individual} $\gamma$-ray strength function, accurately predicts the experimental (p,$\gamma$) reaction data but notably overpredicts at lower energy data points (within the Gamow window for a stellar temperature of 3T$_9$, as indicated). For the S-factor (see Fig~\ref{fig10}), a notable discrepancy exists between the theoretical predictions with these nuclear parameters and the experimental data. However, predictions using the calculated p-OMP, combined with the Skyrme-Hartree-Fock-Bogolyubov level density and BAL $\gamma$-SF, show a better agreement with the present measurements.

In Hauser-Feshbach statistical model calculations, cross-sections are determined based on transmission coefficients, which depend on various width parameters, including particle and gamma widths. These transmission coefficients are derived by solving the Schrödinger equation with specific particle interaction potentials, $\gamma$-ray strength functions, and nuclear level densities. To understand discrepancies between theoretical predictions and experimental results, it is crucial to examine how sensitive different width parameters are, as this can significantly impact the accuracy of cross-section predictions.

Cross-section sensitivity due to width of input parameters (particle and gamma width) can be written as \cite{rauscher2012formalism},    

\begin{equation}
\centering
S_{\sigma i} = \frac{\frac{d\sigma}{\sigma}}{\frac{dW_i}{W_i}}
\end{equation}

\begin{figure}[h]
\begin{center}
\includegraphics[scale=0.3]{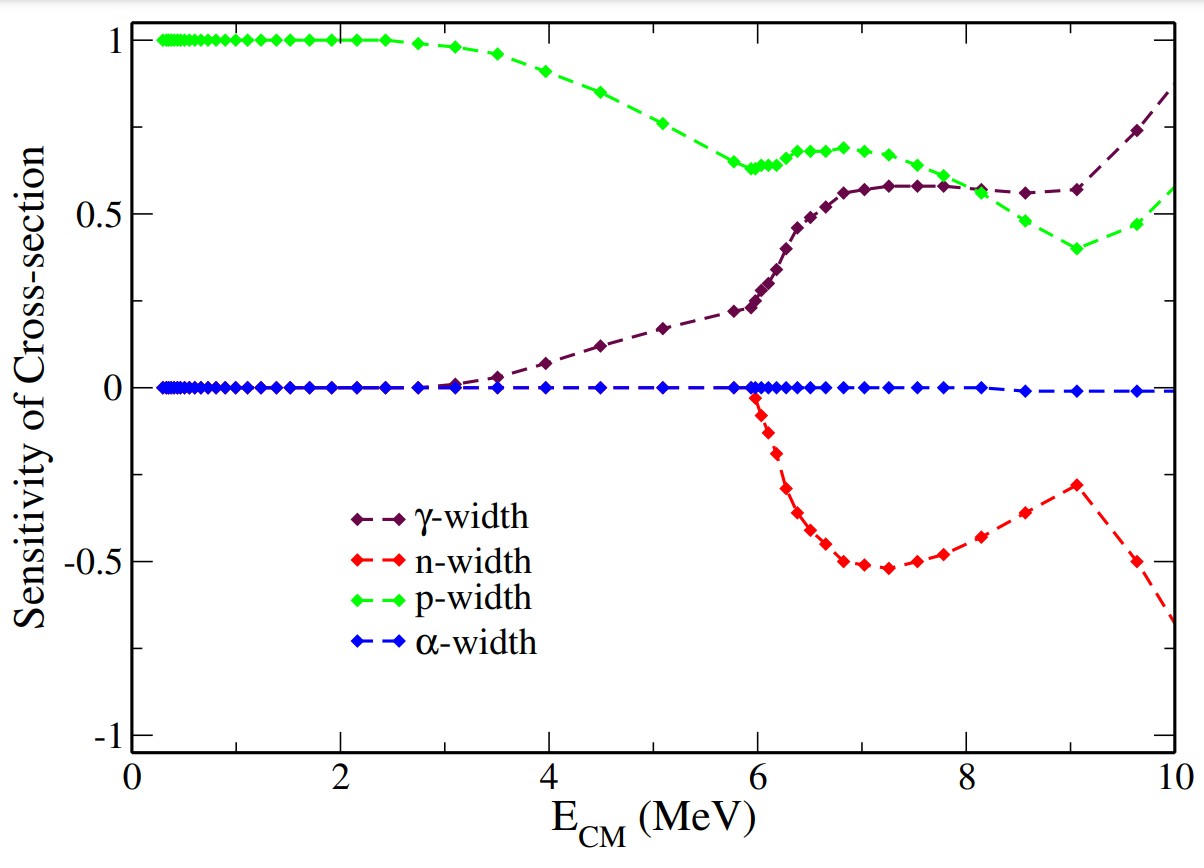}
\end{center}
\caption{The $^{108}$Cd(p,$\gamma$) reaction cross-section sensitivity with centre of mass energy by varying $\alpha, \gamma$, n, p width. Data taken from \cite{rauscher2012formalism}.}\label{fig9}
\end{figure}

In this context, d$\sigma$ and dW$_i$ represent the variations in cross-section and the width of the 
\textit{i}-th input parameter, where \textit{i} can refer to protons, neutrons, alpha particles, or ${\gamma}$-rays. The parameter \textbf{$S_{\sigma i}$} = $\pm$1 indicates that the cross-section changes with the width variation, while \textbf{$S_{\sigma i}$} = 0 shows that the cross-section remains unaffected by changes in width \cite{rauscher2012sensitivity}. Figure~\ref{fig9} illustrates the cross-section sensitivity of the (p,$\gamma$) reaction in relation to different input parameters, highlighting that proton width is particularly sensitive at lower energy levels.

Proton elastic scattering data for 17 different isotopes—specifically p-nuclei and isotopes with mass numbers close to p-nuclei—were obtained from the EXFOR, NNDC database \cite{IAEA_EXFOR, saha2024wood} in January 2023. These data were then analyzed using the optical parameter search code SFRESCO \cite{thompson1988getting, basak2024study}. The calculations focused on proton scattering at the lowest energy range, between 10 and 24.6 MeV. A detailed investigation process is provided in reference \cite{saha2024wood}.
The mass--energy dependent proton optical potential, formulated in the Wood-Saxon form factor,

\begin{eqnarray*}
V_{v} &=& 19.38 + 7.24 \times A^{\frac{1}{3}} - 0.43 \times E + 40.1 \times \frac{N-Z}{A} \\
r_{v} &=& 1.01, \qquad a_{v} = 0.67 \\
W_{s} &=& 8.55 + 0.28 \times \frac{A}{N-Z} \\
r_{s} &=& 1.00, \qquad a_{s} = 0.29 + 3.03 \times \frac{N-Z}{A} \\
V_{SO} &=& 6.36, \qquad r_{SO} = 1.00, \qquad a_{SO} = 0.69
\end{eqnarray*}

Where, \textit{ V$_{v}$, r$_{v}$} and \textit{a$_{v}$ }represent the real volume, radius, and diffusivity terms, while \textit{W$_{s}$, r$_{s}$} and \textit{a$_{s}$ }denote the imaginary surface term, imaginary radius, and diffusivity factor. Additionally, \textit{V$_{SO}$, r$_{SO}$, a$_{SO}$} signify the spin-orbit interaction terms due to the non-zero spin interactions between the projectile and target \cite{ICTP_TALYS_Workshop}.

The potential was derived from parameters fitted to elastic scattering data. When this potential is applied within the TALYS input parameters, it shows a good agreement with the S-factor and cross-section data at lower energy points.

\subsection{S-factor}

\begin{figure}[h]
\begin{center}
\includegraphics[scale=0.3]{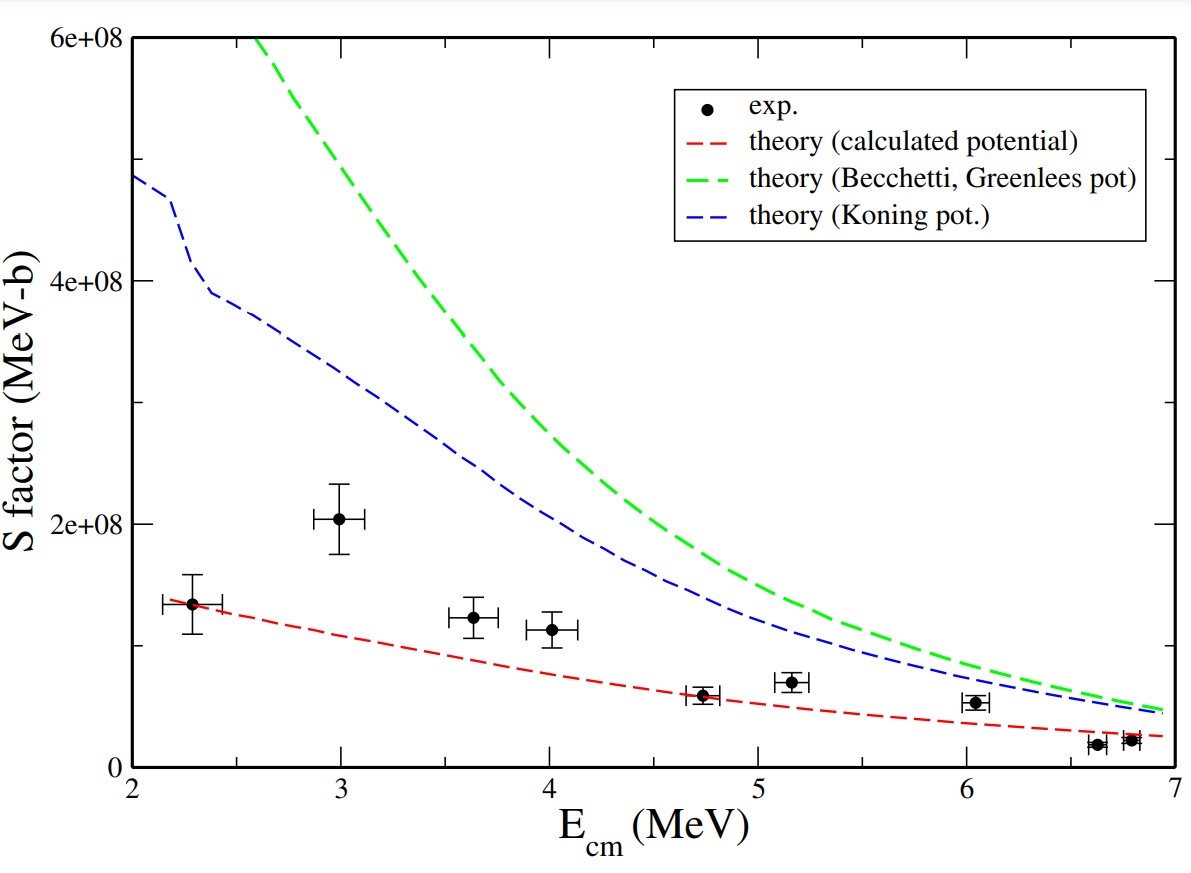}
\end{center}
\caption{Astrophysical S-factor of $^{108}$Cd(p,$\gamma$), theoretically fitted with different input parameter models (Koning potential \cite{koning2003local}, SHFB, BAL; Becchetti et. al potential \cite{becchetti1969nucleon}, SHFB, BAL and Calculated potential, SHFB, BAL) using TALYS-1.96}\label{fig10}
\end{figure}

The astrophysical S-factor is calculated using the given relation \cite{thompson2009nuclear, iliadis2015nuclear}:

\begin{equation}
    S(E_{CM})=E_{CM} \sigma(E_{CM}) e^{2 \pi \eta}
\end{equation}

Where, E$_{CM}$ is the center of mass energy, $\sigma (E_{CM})$ represents the measured cross-section, and $\eta =\frac{Z_1Z_2e^2}{\hbar v}$ is the Sommerfeld parameter. Here, $\textit{v}=\sqrt{\frac{2E}{\mu}}$ denotes the relative velocity, $\mu$ is the reduced mass of the target-projectile system, and $\hbar$ is the reduced Planck constant. The calculated S-factor values are listed in Table~\ref{tab2} and plotted in Figure~\ref{fig10}.

\subsection{Thermonuclear reaction rate}

\begin{figure}[h!]
\begin{center}
\includegraphics[scale=0.3]{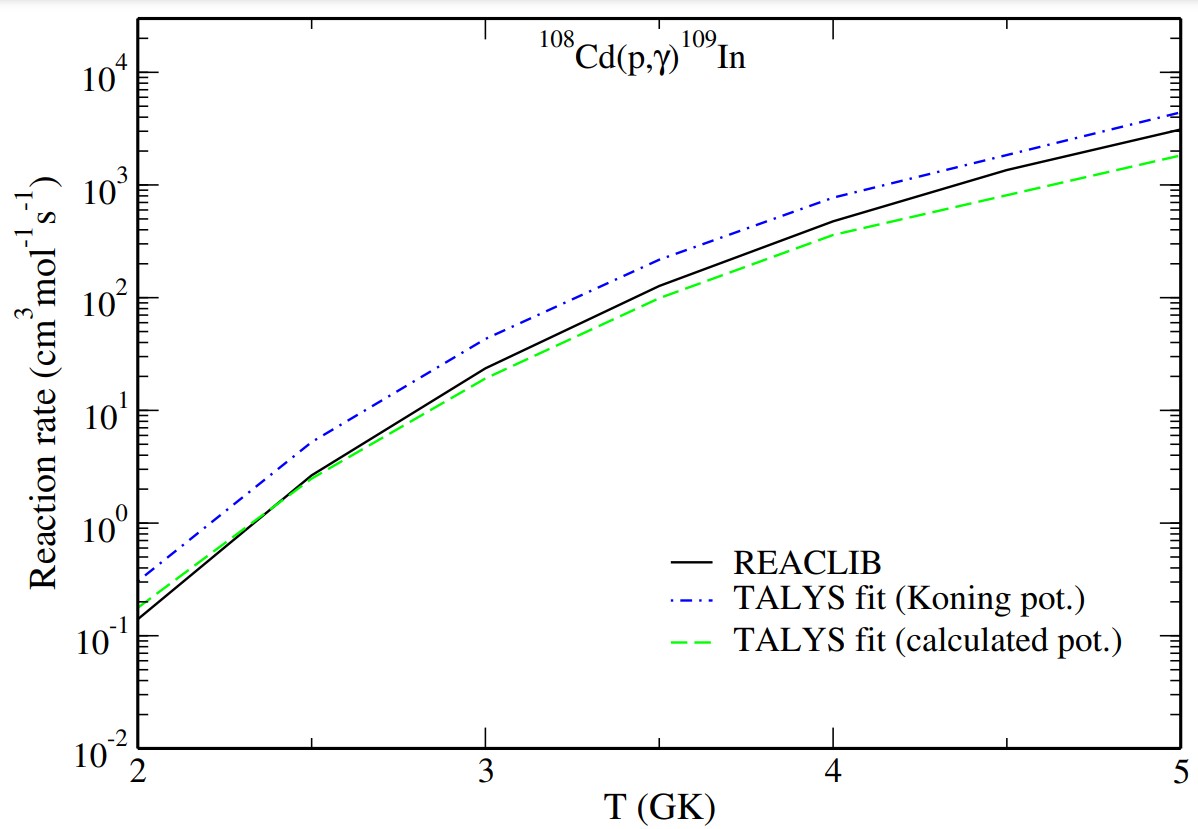}
\end{center}
\caption{Thermonuclear reaction rate of $^{108}$Cd(p,$\gamma$) with stellar temperature (GK) using best fit nuclear input parameter in TALYS-1.96 (Koning potential, SHFB, BAL and Calculated potential, SHFB, BAL) and compared with JINA REACLIB database}\label{fig11}
\end{figure}

\begin{figure}[h!]
\begin{center}
\includegraphics[scale=0.3]{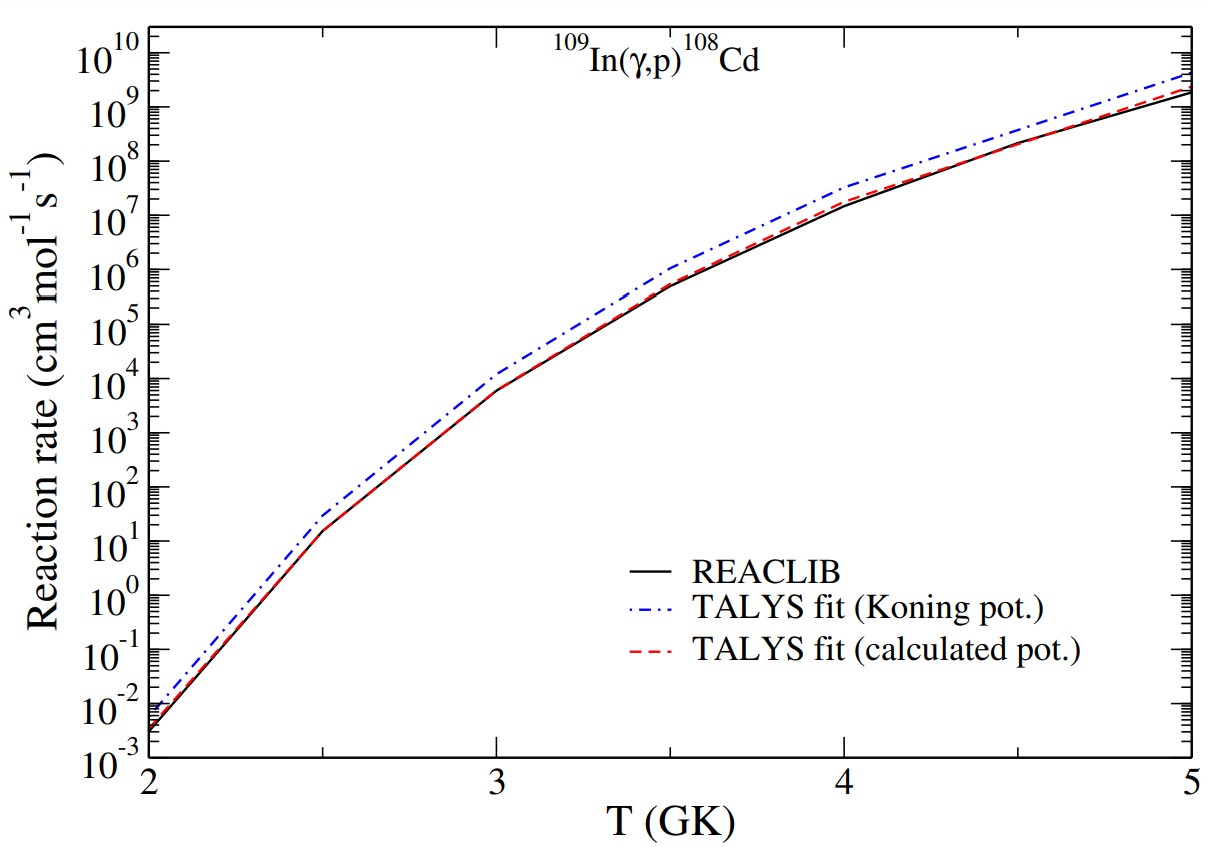}
\end{center}
\caption{Thermonuclear reaction rate of $^{109}$In($\gamma$,p) with stellar temperature (GK) using best fit nuclear input parameter in TALYS-1.96 (Koning potential, SHFB, BAL and Calculated potential, SHFB, BAL) and compared with JINA REACLIB database}\label{fig12}
\end{figure}

The astrophysical reaction rates for 
$^{108}$Cd(p,$\gamma$) and $^{109}$In($\gamma$,p) over a temperature range of 2-5 GK were calculated using TALYS-1.96, as illustrated in Figure~\ref{fig11} and ~\ref{fig12} \cite{rauscher2010relevant}. For these calculations, the nuclear input parameters that provided the best fit were selected, namely the Koning potential combined with the Skyrme-Hartree-Fock-Bogoluybov (SHFB) model for nuclear level density (NLD) and the Brink-Axel-Lorentzian (BAL) model for $\gamma$-ray strength function ($\gamma$-SF), as well as an alternative combination using a calculated potential with the same NLD and $\gamma$-SF models. These reaction rate estimates were then compared to data from the JINA REACLIB database \cite{JINA_ReacLib, cyburt2010jina}, with values listed in Table~\ref{tab3}. The stellar photodissociation rate for $^{109}$In($\gamma$,p) was derived from the capture reaction rate $(N_A \langle \sigma v \rangle_{p,\gamma})$ \cite{rolfs1988cauldrons} for $^{108}$Cd(p,$\gamma$) by applying the reciprocity theorem \cite{arnould2003p, holmes1976tables}. The inverse reaction rate can be expressed as,

\begin{equation}
R _{\gamma,p}(T) = \frac{(2J^o_{^{108}Cd} + 1)(2J_p + 1)}{(2J^o_{^{109}In} + 1)} \cdot \frac{G^o_{^{108}Cd}}{G^o_{^{109}In}} \left( \frac{A_{^{108}Cd} A_p}{A_{^{109}In}} \right)^{3/2} \left(\frac{kT}{2\pi\hbar^2N_A} \right)^{3/2} N_A \langle \sigma v \rangle_{p,\gamma} e^\frac{-Q_{p,\gamma}}{kT}
\end{equation}

Here, \textit{$Q_{p,\gamma}$} represents the Q-value of the $^{108}$Cd(p,$\gamma$) reaction, \textit{k} is Boltzmann’s constant, \textit{$N_A$} is Avogadro's number, \textit{$G^o_i$}(T) is the temperature-dependent normalization partition function, and \textit{$J^o_i$}(T) is the ground state spin of the nucleus labeled `i' with atomic mass number $A_i$.

\begin{table}[h!]
  \centering
  \caption{Astrophysical reaction rate for $^{108}$Cd(p,$\gamma$)$^{109}$In with stellar temperature (GK). Reaction rates are calculated in TALYS-1.96. Fit1 corresponds to the best fitted experimental cross-section data using input parameter TALYS default potential with Skyrme-Hartree-Fock-Bogoluybov (microscopic model) for NLD and Brink-Axel Lorentzian for $\gamma$-SF. Calculated potential is included in Fit2 instead of TALYS default potential.}\label{tab3}
   \begin{tabular}{llll} 
    \hline
        \textbf{Stellar} & \textbf{Rate (fit1)} &  \textbf{Rate (fit2)} & \textbf{REACLIB data} \\
        \textbf{temp. (GK)} & \textbf{(cm$^3$ s$^{-1}$ mol$^{-1}$)} & \textbf {(cm$^3$ s$^{-1}$ mol$^{-1}$)} & \textbf{(cm$^3$ s$^{-1}$ mol$^{-1}$)} \\
     \hline   
    
      2 & 0.30 & 0.18 & 0.14 \\
      2.5 & 5.23 & 2.48 & 2.65 \\
      3 & 43.0 & 19.2 & 23.6 \\
      3.5 & 217 & 99 & 127 \\
      4 & 772 & 360 & 476 \\
     \hline  
\end{tabular}
\end{table}

\begin{table}[h!]
  \centering
  \caption{Same as table 3, but for the $^{109}$In($\gamma$,p)$^{108}$Cd reaction.}\label{tbl4}
   \begin{tabular}{llll} 
    \hline
        \textbf{Stellar} & \textbf{Rate (fit1)} &  \textbf{Rate (fit2)} & \textbf{REACLIB data} \\
        \textbf{temp. (GK)} & \textbf{(cm$^3$ s$^{-1}$ mol$^{-1}$)} & \textbf {(cm$^3$ s$^{-1}$ mol$^{-1}$)} & \textbf{(cm$^3$ s$^{-1}$ mol$^{-1}$)} \\
     \hline   
    
      2 & 0.005 & 0.003 & 0.003 \\
      2.5 & 29.6 & 15.3 & 15.36 \\
      3 & 12000 & 6050 & 5960 \\
      3.5 & 1.05$\times 10^{6}$ & 5.42$\times 10^{5}$ & 4.93$\times 10^{5}$ \\
      4 & 3.27$\times 10^{7}$ & 1.77$\times 10^{7}$ & 1.47$\times 10^{7}$ \\
     \hline  
\end{tabular}
\end{table}

\section{Conclusion}

For p-process modeling, it is crucial to have precise knowledge of the reaction rates for hundreds of nuclear reactions within the reaction networks. In this study, the cross-section for the $^{108}$Cd(p,$\gamma$) reaction was measured using the stack foil activation method, with laboratory energies ranging from \textbf{2.26$\pm$0.12} MeV to 
\textbf{6.84$\pm$0.03} MeV, which are relevant to p-process nucleosynthesis. This is the first report of the reaction cross-section at \textbf{2.26$\pm$0.12} MeV, marking it as the lowest energy point within the Gamow window for a temperature of 3 GK. The S-factor and reaction rates were derived from the measured data.

The experimental results were compared to theoretical predictions made using the TALYS-1.96 statistical model code, where various input parameters were adjusted. A newly calculated optical potential for p-nuclei, along with Skyrme-Hartree-Fock-Bogoliubov level densities and Brink-Axel-Lorentzian $\gamma$-ray strength functions, was implemented in TALYS-1.96 to achieve better alignment with the lower energy cross-sections and S-factor data points. The optimized input parameters were then employed to calculate the reaction rates at temperatures ranging from 2 to 4 GK. Additionally, the $^{109}$In($\gamma$,p) reaction rate was computed for the $^{108}$Cd(p,$\gamma$) reaction data using TALYS-1.96 and the reciprocity theorem, revealing that the reaction rate data differs from the LUNA REACLIB values by a factor of less than 1.5.

\section*{Acknowledgement}

The authors extend their gratitude to Dr. Chandana Bhattacharya and Dr. Vaishali Naik for their invaluable support in the successful completion of the experiment. Special thanks are given to Mr. A. A. Mallick from the Analytical Chemistry Division at VECC, as well as the team members of the K130 Cyclotron Facility at VECC, Kolkata. The authors also acknowledge Prof. Supratic Chakraborty and Mr. Gautam Sarkar from the Surface Physics and Material Science Division, SINP, Kolkata for their assistance in target preparation and characterization. Sukhendu Saha acknowledges the Council of Scientific and Industrial Research (CSIR), Government of India, for funding support (File No.: 09/489(0119)/2019-EMR-I).

\section*{References}

\bibliography{108cd_jpg}

\end{document}